\documentclass[namedreferences]{SolarPhysics}
\usepackage[optionalrh]{spr-sola-addons} % For Solar Physics 
\usepackage{graphicx}                    % For eps figures, newer & more powerfull
\usepackage{color}                       % For color text: \color command
\usepackage{url}                         % For breaking URLs easily trough lines
\newcommand{\solphys}{{\it Solar Phys.}}

\begin{document}

\begin{article}

\begin{opening}

\title{Determination of Electromagnetic Source Direction as an Eigenvalue Problem}

%%%%%%%%%%%%%%%%%%%%%%%%%%%%%%%%%%%%%%%%%%%%%%%%%%%
%% Authors Names
%
\author{Juan C.~\surname{Mart\'inez-Oliveros}$^{1}$\sep
        Charles~\surname{Lindsey}$^{2}$\sep\\
        Stuart D.~\surname{Bale}$^{1, 3}$\sep
        S\"am~\surname{Krucker}$^{1,4}$
       }       

%%%%%%%%%%%%%%%%%%%%%%%%%%%%%%%%%%%%%%%%%%%%%%%%%%%
%% Runningheads
%
\runningauthor{J.C. Mart\'{\i}nez-Oliveros \textit{et al.}}
\runningtitle{Determination of Electromagnetic Source Direction as an Eigenvalue Problem}

%%%%%%%%%%%%%%%%%%%%%%%%%%%%%%%%%%%%%%%%%%%%%%%%%%%
%% Affilations 
%
  \institute{$^{1}$ Juan C. Mart\'inez-Oliveros  \sep Stuart D. Bale \sep S\"am Krucker \\ Space Science Laboratory, University of California, Berkeley, CA 94720, USA \\
                   email: \url{oliveros@ssl.berkeley.edu}, \url{bale@ssl.berkeley.edu}, \url{krucker@ssl.berkeley.edu} \\ 
  		$^{2}$ Charles A. Lindsey\\ Colorado Research Associates, NWRA, Boulder, CO, USA\\
		email: \url{clindsey@cora.nwra.com}\\
		$^{3}$ Stuart D. Bale\\Physics Department, University of California, Berkeley, CA  94720, USA\\
		$^{4}$ S\"am Krucker\\Institute of 4D Technologies, School of Engineering, University of Applied Sciences North Western Switzerland, 5210 Windisch, Switzerland}

%%%%%%%%%%%%%%%%%%%%%%%%%%%%%%%%%%%%%%%%%%%%%%%%%%%
%%% Abstract 
\begin{abstract}
Low-frequency solar and interplanetary radio bursts are generated at frequencies below the ionospheric plasma cutoff and must therefore be measured in space, with deployable antenna systems.  The problem of measuring both the general direction and polarization of an electromagnetic source is commonly solved by iterative fitting methods such as linear regression that deal simultaneously with both directional and polarization parameters. We have developed a scheme that separates the problem of deriving the source direction from that of determining the polarization, avoiding iteration in a multi-dimensional manifold. The crux of the method is to first determine the source direction independently of concerns as to its polarization. Once the source direction is known, its direct characterization in terms of Stokes vectors in a single iteration if desired, is relatively simple.  This study applies the source-direction determination to radio signatures of flares received by STEREO. We studied two previously analyzed radio type III bursts and found that the results of the eigenvalue decomposition technique are consistent with those obtained previously by \citeauthor{2009SoPh..259..255R} (\solphys{} \textbf{259}, 255, \citeyear{2009SoPh..259..255R}). For the type III burst observed on 7 December 2007, the difference in travel times from the derived source location to STEREO A and B is the same as the difference in the onset times of the burst profiles measured by the two spacecraft. This is consistent with emission originating from a single, relatively compact source. For the second event of 29 January 2008, the relative timing does not agree, suggesting emission from two sources separated by 0.1 AU, or perhaps from an elongated region encompassing the apparent source locations.

\end{abstract}

%%%%%%%%%%%%%%%%%%%%%%%%%%%%%%%%%%%%%%%%%%%%%%%%%%%
%% Keywords
%
\keywords{Solar wind, Direction finding, Type III bursts, Solar radio}

\end{opening}

%-------------------------------------------------

\section{Introduction}
     \label{S-Introduction} 

Solar activity encompasses a wide range of phenomena that may vary in complexity. Solar flares and coronal mass ejections (CMEs) are among the best known and most spectacular examples of this activity. During flares and CMEs, energy stored in the coronal magnetic field is converted into kinetic energy and thermal energy which are observed as coronal mass movement, and used to heat the plasma adjacent to the reconnection region, and in the acceleration of particles that eventually will follow magnetic field lines in the interplanetary medium.

Flares and CMEs emit radiation at a variety of frequencies, from X-rays at the coronal sources to metric and decametric radio emission in the interplanetary medium. Each observed frequency range can then be associated with a specific mechanism of emission, which gives clues about the physical processes taking place in the solar atmosphere or interplanetary medium when the emission emanated. 

The radio emission observed at metric and decametric wavelengths consists of a long series of ``radio bursts'' that may last from a period of hours to days \cite{1998ARA&A..36..131B}.  The physical processes at work, including the changing ambient conditions, are signified by temporal evolution in the frequency spectrum of the emission.  Historically, solar radio bursts have been classified into types, based on the morphology of features that appear in plots of electromagnetic power in a frequency-time plane. Type II radio bursts are generally observed during major flares and/or CMEs associated with filament eruptions, at meter wavelengths for periods from 5 to 30 min. These bursts show a distinctive frequency drift in time, from higher frequencies to lower frequencies. Type III bursts are the most common, believed to be the result of flare-energized electrons that produce ``plasma emission'' in the meter and decimeter range with lifetimes of approximately 10 s. In general, type III events appear in groups lasting several minutes, which indicates that each observed event is a collection of many small-scale type III bursts. In this paper we are interested in interplanetary radio emissions from type III bursts, particularly the direction of the radio source, but the results can also be applied to type II bursts. 

Radio direction finding (DF) is a powerful technique that allows us to determine the direction of a radio source, using observations made with simple, deployable electric antennas. There are several methods to do this, depending on how the radio emission is received and recorded.  Some methods require distinguishing signal modulation in receivers on spinning spacecraft \cite{1972Sci...178..743F}.  Others are based on auto- and cross-correlation analyses of data from three-axis-stabilized spacecraft, such as on NASA's \textit{Solar Terrestrial Relations Observatory} (STEREO) \cite{2008SSRv..136....5K,2008SSRv..136..487B}, \textit{Cassini}, and future missions like \textit{Solar Probe Plus} and \textit{Solar Orbiter}.   Analysis techniques for these data sets can be divided into two groups:  iterative fitting \cite{1995RaSc...30.1699L,2004JGRA..10909S17V,2009SoPh..259..255R} and noniterative methods \cite{santolik2003,2005RaSc...40S3003C,2010RaSc...45S3003H}. To fit observational data, the methods in the first group use theoretical models of the auto- and cross-correlations to find the direction and polarization of the radio emission \cite{1995RaSc...30.1699L,2005RaSc...40S3003C}. The method used by  \inlinecite{2009SoPh..259..255R} and \inlinecite{2004JGRA..10909S17V} uses a least-squares minimization fitting of the data, while the method described in \inlinecite{1995RaSc...30.1699L} iterates the solution obtained using a singular value decomposition (SVD) process. The \inlinecite{1995RaSc...30.1699L} method is a more sophisticated iterative fitting method that allows one to estimate the factor of error of each solution found by the iteration process. 

Noniterative methods encompass different techniques to solve the direction finding problem. For most polarizations, it is possible to find the solution by inverting the particular case of the general solution of the auto- and cross-correlations, as is shown in \inlinecite{2005RaSc...40S3003C}. This method gives an efficient and fast solution to the problem; however, its  downside is that it constrains the polarization of the radio signal, which is generally unknown. \inlinecite{santolik2003} implemented a method that locates the direction of emission of electromagnetic waves, by finding the singular values of the spectral matrix, a $3 \times 6$ array composed of the values of the auto- and cross-correlations, noting that the values of the cross-correlations are generally complex.  The relation between the singular values obtained using an SVD algorithm gives the direction of the radio source and certain properties of the polarization.

All the foregoing examples address the problem of simultaneous determination of both the source direction and its polarization, the latter in terms of Stokes vectors.  In this paper we develop a simplified method to derive the general direction of the radio source distribution independently from its polarization. This method is similar in form to the SVD method proposed by \inlinecite{santolik2003}, but it addresses only the problem of finding the location of the radio source, using an eigenvalue decomposition method.  We apply the method to two type III bursts analyzed by \inlinecite{2009SoPh..259..255R} and compare our results with theirs.

\section{The Eigenvalue Decomposition Method}
\label{S-EDM} 

We will suppose that the wave electric field $\mathbf{E}$ in the vicinity of the spacecraft can be expressed at a specific time by
\begin{equation} \label{eq1}
\mathbf{E}^\textrm{r} (t) = {\mathrm{Re}}\{\mathbf{E}e^{i\omega t}\} ,
\end{equation}
with the harmonic wave frequency, $\omega$,  assumed constant, and with $\mathbf{E}$ generally complex. We first consider the case in which $\mathbf{E}$ is constant, noting the following: 

\begin{enumerate}
\item $\mathbf{E}^\textrm{r} (t)$ moves in a plane, $\mathcal{P}$, generally tracing out an ellipse. 
\item If $\mathbf{E}^\textrm{r} (t)$ is the result of a single point source, then the direction of the source from the spacecraft must be along the line, $\mathcal{R}$, perpendicular to $\mathcal{P}$, if $\mathcal{P}$ is uniquely determined. This could be either of both directions from the spacecraft in which $\mathcal{R}$ extends. If the motion of $\mathbf{E}^\textrm{r} (t)$ is confined to some line, $\mathcal{L}$, then there is no unique $\mathcal{P}$, in which case the source can be anywhere in the plane perpendicular to $\mathcal{L}$ in which the spacecraft sits at any moment. 
\end{enumerate}

To prove that $\mathbf{E}^\textrm{r} (t)$ moves in some plane, simply observe from Equation~(\ref{eq1}) that
\begin{equation} \label{eq2}
\mathbf{E}^\mathrm{r} (t) = {\mathrm{Re}}\{\mathbf{E}\} \cos{\omega t} 
 - {\mathrm {Im}}\{\mathbf{E}\} \sin{\omega t}, 
\end{equation}
\textit{i.e.}, $\mathbf{E}^\textrm{r} (t)$ is a linear combination of two vectors ${\rm Re}\{\mathbf{E}\}$ and ${\rm Im}\{\mathbf{E}\}$. If ${\rm Re}\{\mathbf{E}\}$ and ${\rm Im}\{\mathbf{E}\}$ are linearly independent, then they uniquely define a plane, $\mathcal{P}$, and $\mathbf{E}^\textrm{r} (t)$ must be contained in that plane. Otherwise, if ${\rm Re}\{\mathbf{E}\}$ and ${\rm Im}\{\mathbf{E}\}$ are parallel, $\mathbf{E}^\textrm{r} (t)$ is confined to a linear motion, which is contained in an infinite number of planes. Either way, motion in at least one plane is secured, but if ${\rm Re}\{\mathbf{E}\}$ and ${\rm Im}\{\mathbf{E}\}$ are linearly independent, then the plane, $\mathcal{P}$, in which $\mathbf{E}^\textrm{r}$ moves is unique, whereas if they are parallel, then $\mathcal{P}$ is indefinite, and the direction to the source accordingly ambiguous. 

We defer the proof that the motion of $\mathbf{E}^\textrm{r}$ in $plane \mathcal{P}$ is elliptical, as a determination of source direction does not depend on it. 

The second point above is a result of well-established classical radiation theory, described at length in most texts. It applies to any source, not just a monochromatic one such as Equation~(\ref{eq1}) expresses when $\mathbf{E}$ is constant. 

When ${\rm Re}\{\mathbf{E}\}$ and ${\rm Im}\{\mathbf{E}\}$ are linearly independent, the source direction in the case of an electric field, $\mathbf{E}$, known to be from a single source can be expressed by the unit vector, $\hat{\mathbf{R}}$, in the direction of 
\begin{equation} \label{eq3}
\mathbf{R} = {\rm Re}\{\mathbf{E}\} \times {\rm Im}\{\mathbf{E}\}, 
\end{equation}
meaning that the source is either in the direction of $\hat{\mathbf{R}}$ or $-\hat{\mathbf{R}}$. Otherwise (\textit{i.e.}, ${\rm Re}\{\mathbf{E}\}$ and ${\rm Im}\{\mathbf{E}\}$ are parallel), $\mathbf{R}$ is null, in which case the source can be anywhere in the plane perpendicular to either. (Besides the above, there is only the extreme case of a null source, in which both ${\rm Re}\{\mathbf{E}\}$ and ${\rm Im}\{\mathbf{E}\}$ are null. In that case, of course, the source can be anywhere at all.) 

\subsection{The Case of a Distributed Source}

In the case of a completely coherent distributed source, meaning that the resultant $\mathbf{E}$ from all of the source components is temporally invariant throughout the measurement, $\mathbf{E}^\textrm{r} (t)$ would remain confined to move in a plane, as proved above. In practice, far-separated source components tend not to be fully coherent. Consequently, Equation~(\ref{eq1}) can still express the motion of $\mathbf{E}^\textrm{r} (t)$, but the direction, amplitude, and phase of $\mathbf{E}$ are now all subject to some variation, such that $\mathbf{E}^\textrm{r} (t)$ is no longer strictly confined to a plane. For this problem, it is useful to regard $\hat{\mathbf{R}}$ as the direction through the center of the source distribution by some definition. If the bandwidth of the radiation received by the spacecraft is $\Delta\omega$ centered on $\omega$, then the plane in which $\mathbf{E}^\textrm{r} (t)$ moves will itself tend to nutate with respect to $\hat{\mathbf{R}}$ on a time scale of $1/\Delta\omega$. Hence, $\mathbf{E}^\textrm{r} \cdot \hat{\mathbf{R}}$ will no longer be entirely null. 

If the source is relatively compact, then  $\mathbf{E}^\textrm{r} \cdot \hat{\mathbf{R}}$  will always be relatively small. We therefore now define a representative ``source direction," $\hat{\mathbf{R}}$, for a distributed source to be either of the two opposite directions from the spacecraft for which $\langle( \mathbf{E}^\textrm{r} \cdot \hat{\mathbf{R}})^2 \rangle$ is minimal (or infinitely many directions for the various cases in which a unique minimum is nonexistent). Here, the angular brackets signify an average over the time interval of the measurement. The problem, then, reduces to the determination of the line $\mathcal{R}$ on which $\hat{\mathbf{R}}$ lies when $\langle( \mathbf{E}^\textrm{r} \cdot \hat{\mathbf{R}})^2 \rangle$ is uniquely minimal or the plane to which $\hat{\mathbf{R}}$ is confined, if any, when $\langle( \mathbf{E}^\textrm{r} \cdot \hat{\mathbf{R}})^2 \rangle$ is not uniquely minimal. 

The relative magnitude of the minimum of $\langle( \mathbf{E}^\textrm{r} \cdot \hat{\mathbf{R}})^2 \rangle$ with respect to the total power, $\langle(\mathbf{E}^\textrm{r})^2\rangle$, can be regarded as the square of a characteristic radius, $r_\mathrm{c}$, of the source distribution.
This is explained in Appendix \ref{appendix-a}.

\subsection{Observational Data}

In general, spacecraft have a reference frame, $\mathcal{S}$, that can be expressed by a set of orthonormal vectors, 
\begin{equation} \label{eq4}
\mathcal{S} \equiv \{\mathbf{\hat{e}}_1 , \mathbf{\hat{e}}_2, 
\mathbf{\hat{e}}_3\}, 
\end{equation}
pointing in the $x$-, $y$-, and $z$-directions, respectively. Hence, $\mathbf{E}$ can be expressed in terms of complex components, $E_\alpha$, thus: 
\begin{equation} \label{eq5}
\mathbf{E}= \sum_\alpha E_\alpha \hat{\mathbf{e}}_\alpha,
\end{equation}
with
\begin{equation} \label{eq6}
\alpha =\{1,2,3\}.
\end{equation}
For STEREO, the products used to determine the position of the radio source are complex statistical correlations between electric-vector components for some frequency, $\omega$, comprising a $\mathrm{3 \times 3}$ matrix, $\mathbf{C}$, whose elements are 
\begin{equation} \label{eq7}
C_{\alpha \beta} \equiv \langle E_\alpha E^{*}_\beta \rangle,
\end{equation}
with $\alpha$, $\beta \in \{1, 2, 3\}$, in some bandwidth, $\Delta \omega$, about $\omega$, where $\mathbf{E}$ is defined by Equation~(\ref{eq1}) and the angular brackets indicate an average over a time interval, $\Delta t$. 

\subsubsection{Interpretation of the Correlations}

It is fairly straightforward to determine $\langle( \mathbf{E}^\textrm{r} \cdot \hat{\mathbf{R}})^2 \rangle$ in any direction from $\mathbf{C}$ in any reference frame. We first define the real correlation, $\mathbf{C}^\textrm{r}$ , by 
\begin{equation} \label{eq8}
C^\textit{r}_{\alpha \beta} = \langle E^\textrm{r}_\alpha(t) E^\textrm{r}_\beta(t) \rangle , 
\end{equation}
where the angular brackets indicate averaging over time, and show that this can be determined from $\mathbf{C}$: 
\begin{eqnarray} \label{eq9}
C^\textit{r}_{\alpha \beta} &=& \langle {\rm Re}\{E_\alpha e^{i\omega t}\}    {\rm Re}\{E_\beta e^{i\omega t}\}  \rangle \nonumber \\ 
 & =& \frac{1}{4} \langle(E_\alpha e^{i\omega t} +E^*_\alpha e^{-i\omega t}) (E_\beta e^{i\omega t} +E^*_\beta e^{-i\omega t})  \rangle .
\end{eqnarray}
At this point, we will apply the distributive property to the product on the right side of Equation~(\ref{eq9}), noting that this will have terms containing $\exp(in\omega t)$ with $n \in \{0, \pm2\}$. We now assume that the spectrum of $\mathbf{E}$ outside of a bandwidth $\Delta \omega < \omega/2$ is either null or negligible. Thus, we can drop all terms for which $n \in \{\pm 2\}$, since the averages of these will be null (or similarly negligible). This leaves us with only the cross terms in which an occurrence of  $\exp(i\omega t) $ is offset by its inverse,  $\exp(-i\omega t)$, hence, 
\begin{eqnarray} \label{eq10}
C^\textit{r}_{\alpha \beta} &=& \frac{1}{4}\langle E_\alpha E^*_\beta  +E_\beta E^*_\alpha \rangle \nonumber \\ 
 & =& \frac{1}{2} {\rm Re}\{\langle E_\alpha E^*_\beta \rangle \} \nonumber \\
 & =& \frac{1}{2} {\rm Re}\{C_{\alpha \beta} \} .
\end{eqnarray}

\noindent Stated slightly more succinctly, 
\begin{equation}\label{eq11}
\mathbf{C}^\textrm{r} =  \frac{1}{2} {\rm Re}\{\mathbf{C} \}.
\end{equation}
Thus, the correlations between the real electric components averaged over time are simply half of the real parts of the complex correlations. 

Note that the diagonal components of $\mathbf{C}^\textrm{r}$ are the mean squares of the electric components, $\langle( \mathbf{E}^\textrm{r} \cdot {\mathbf{\hat{e}}_\alpha})^2 \rangle$, in the frame of the spacecraft. Now, given the real 
matrix, ${\bf C}^\textit{r}$, in any orthonormal coordinate system, $\cal S$, it is straightforward to determine its analog, ${{\bf C}^\textit{r}}'$, in any other orthonormal frame,

\begin{equation} \label{eq12}
{\mathcal S}' ~\equiv~ \{{\mathbf{\hat{e}}}_1', ~{\mathbf{\hat{e}}}_2', ~{\mathbf{\hat{e}}}_3'\},
\end{equation} %\eqno{(15)}
from its diagonal components $\langle({\bf E}^\textit{r}\cdot{\bf\mathbf e}_\alpha')^2\rangle$, in ${\mathcal S}'$. Hence, knowing $\mathbf{C}^\textrm{r}$ in any single frame, $\mathcal{S}$, allows us to determine $\mathbf{C}^\textrm{r}{'}$ in any other frame, $\mathcal{S}{'}$, thus to take from its diagonal components $\langle( \mathbf{E}^\textrm{r} \cdot {\mathbf{\hat{e}}{'}_\alpha})^2 \rangle$, in $\mathcal{S}{'}$. 

To see that this is true, let $\mathbf{U}$ represent the orthonormal transformation from $\mathcal{S}$ to $\mathcal{S}{'}$. To be particular, we express $\mathbf{U}$ by the real, orthonormal matrix whose elements are $U_{\alpha \beta}$, with $\alpha$, $\beta \in \{1, 2, 3\}$, in that the components, $E{'}_\alpha =  \mathbf{E} \cdot \mathbf{\hat{e}}{'}_\alpha$, of $\mathbf{E}$ in  $\mathcal{S}{'}$ are 
\begin{equation} \label{eq13}
E^\textrm{r}_\alpha{'} = \sum_\mu U_{\alpha \mu} E^\textrm{r}_\mu.
\end{equation}

\noindent Extending Equation (8) to the transformed reference frame, it now follows that
\begin{eqnarray} \label{eq14}
{C_{\alpha\beta}^\textit{r}}'  &=&  \langle {E_{\alpha}^\textit{r}}' {E_{\beta}^\textit{r}}'\rangle \nonumber \\
 &=& \Big\langle \sum_\mu U_{\alpha\mu} E_{\mu}^\textit{r} \sum_\nu U_{\beta\nu}  E_{\nu}^\textit{r}\Big\rangle \nonumber \\
 &=& \sum_{\mu\nu} U_{\alpha\mu} \langle E_{\mu}^\textit{r}   E_{\nu}^\textit{r}\rangle U_{\nu\beta}^\mathrm{T} \nonumber \\
 &=& \sum_{\mu\nu} U_{\alpha\mu} C_{\mu\nu}^\textit{r} U_{\nu\beta}^\mathrm{T},
\end{eqnarray}

\noindent where $U_{\nu\beta}^\mathrm{T} \equiv U_{\beta\nu}$ defines the transpose of ${\bf U}$, and because ${\bf U}$ is real and orthogonal, ${\bf U}^\mathrm{T}$ is the inverse of ${\bf U}$. Hence, 
\begin{equation} \label{eq15}
\mathbf{C}^\textrm{r}{'} = \mathbf{U} \mathbf{C}^\textrm{r} \mathbf{U}^{-1}.
\end{equation}
Note that, because multiplication is commutative,
\begin{equation} \label{eq16}
\langle E^\textrm{r}_\alpha E^\textrm{r}_\beta \rangle = \langle E^\textrm{r}_\beta E^\textrm{r}_\alpha \rangle,
\end{equation}
and it therefore follows that
\begin{equation} \label{eq17}
C^\textit{r}_{\alpha \beta} = C^\textit{r}_{\beta \alpha},
\end{equation}
\textit{i.e.}, $\mathbf{C}^\textrm{r}$ is symmetric. Thus, there exists an orthonormal rotation, $\mathbf{U}$, that diagonalizes $\mathbf{C}^\textrm{r}$. This is simply the matrix whose columns are the eigenvectors of $\mathbf{C}^\textrm{r}$.

Since $\mathbf{C}^\textrm{r}$ is real and symmetric, its eigenvalues must be real. Indeed, since the diagonal components of $E^\textrm{r}_\alpha{'}$ are correlations of components of $\mathbf{E}^\textrm{r}$ with themselves for whatever $\mathbf{U}$, the eigenvalues must be non-negative. These eigenvalues will generally have a greatest and a least value. Broadly speaking, a three-dimensional map of $\langle (\mathbf{E}^\textrm{r} \cdot \mathbf{\hat{r}})^2 \rangle$ as a function of direction, $\mathbf{\hat{r}}$, traces a three-dimensional ellipsoid whose minor axis is the least eigenvalue of ${C}^\textit{r}$ in length, in the direction of the corresponding eigenvector; whose major axis is the greatest eigenvalue of ${C}^\textit{r}$ in length, in the direction of the eigenvector corresponding to that (perpendicular to the first); and to whose intermediate axis (perpendicular to the other two) is attached an eigenvalue somewhere in between the least and the greatest. The source direction, $\mathbf{\hat{R}}$, is taken to be along the minor axis of the foregoing ellipsoid. 

There are special cases leading to degeneracies in which any two or all three eigenvalues can be equal, or in which one or more can be null (all three for a null source). In those cases, the source direction becomes less definite in certain respects similar to those previously noted in the special case of a point source. In the case of two equal eigenvalues less than $1/3$, we can only suppose that $\mathbf{\hat{R}}$ lies in the plane spanned by the two corresponding eigenvectors. In the extreme case of a null source or, e.g., a completely isotropic distribution of randomly polarized noise, all three eigenvectors are equal, and the source direction is then completely indefinite. When there is a single least eigenvalue, the eigenvector that corresponds to it is the one to be identified with $\mathbf{\hat{R}}$, the direction toward (or away from) the center of the source distribution. 

Thus the problem of finding $\mathbf{\hat{R}}$ reduces to the eigenvalue problem of diagonalizing the real matrix ${\rm Re}\{\mathbf{C}\}/2$, selecting the least of its eigenvalues and identifying the eigenvector that corresponds to that eigenvalue. The crux of the eigenvalue problem is solving for the roots of a secular equation that is generally cubic.  All three roots are ensured to be real and positive. The cubic equation can be solved in any number of ways.

Once the source direction is determined from a minimal eigenvector, one can straightforwardly determine the polarization of the source in terms of Stokes vectors from the directions of the maximal and intermediate eigenvalues (or multiply maximal in the case of one or more equalities) and the eigenvectors attached to them. This is the subject of a continuation of this study.  The relative value of the minimal eigenvector similarly allows us to model the effective angle subtended by the source distribution. 

\begin{figure}[t]
\centering
\includegraphics[width=0.9\columnwidth]{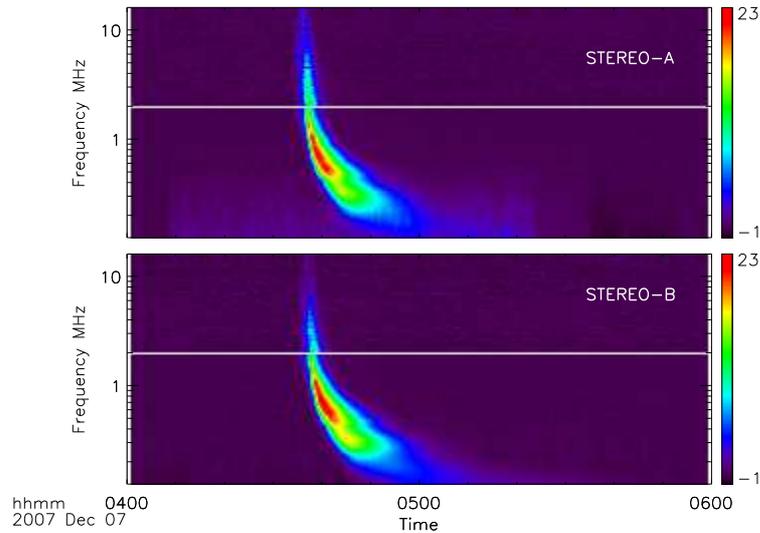}
\caption{STEREO A and  B dynamic spectra of the 7 December 2007 type III burst from 04:00~UT to 06:00~UT. The plotted frequencies range from 100~kHz to 16.025~MHz. The color bars represent the intensity of the radio emission measured in decibels.}
\label{fig1}
\end{figure}

\section{The STEREO Radio Observations}
     \label{S-Observations} 

For application of the formalism to STEREO/WAVES observations, we developed a set of codes that allow us to determine and plot the positions of type II and type III radio sources. In this section we present and compare the results of the method applied to two well-studied type III bursts by \inlinecite{2009SoPh..259..255R}.

\subsection{7 December 2007 Type III Event}
\label{sec31}

A solar flare classified as a GOES~B1.4 was observed on 7 December 2007, hosted by the active region 10977 located close to the disk center (S05W06). The flare started at 04:36~UT, reaching maximum at 04:39~UT and ending at 05:01~UT. This flare was associated with a type III burst registered by space- and ground-based radio observatories, including both STEREO and \textit{Wind} spacecraft, with a starting time of about 04:36~UT (Figure~\ref{fig1}).  

\begin{figure}[p]
\centering
\includegraphics[angle=90, clip,width=1\columnwidth]{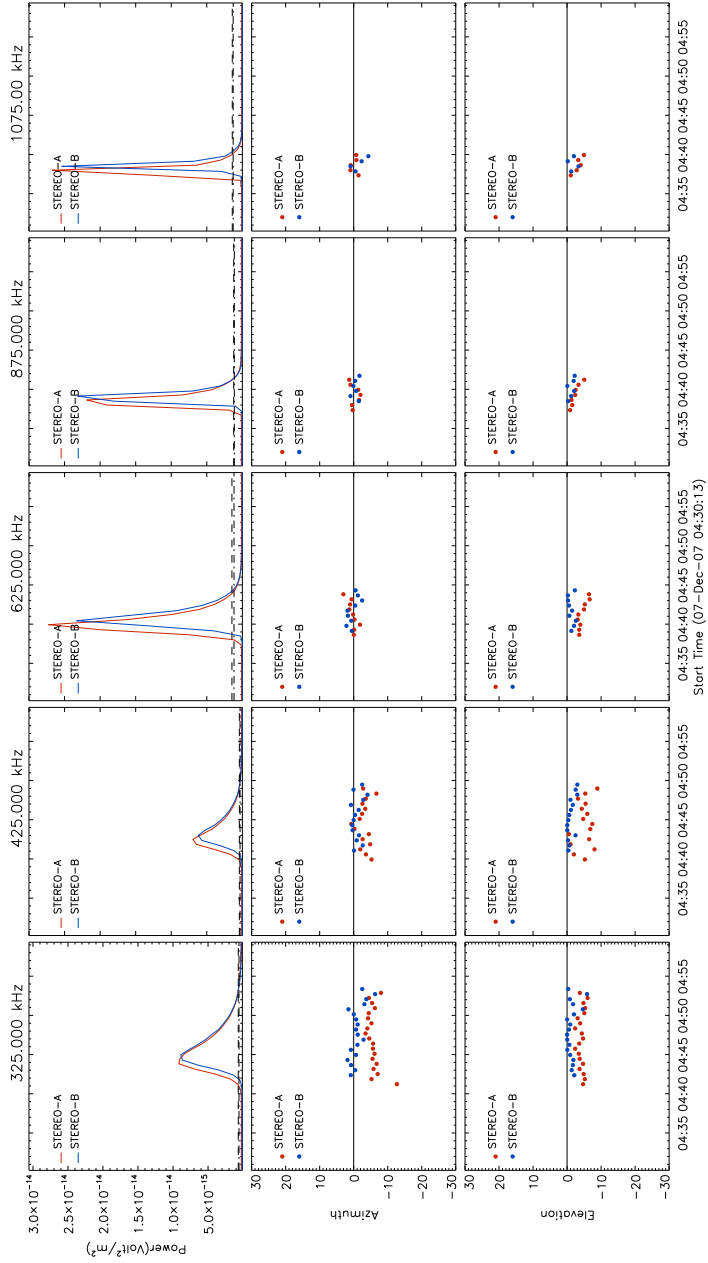}
\caption{Results of the direction finding analysis at STEREO A and B at 325, 425, 625, 875 and 1075~kHz for the 7 December 2007 solar type III burst. The top panels show the derived radio emission power, the middle panels show source azimuths with respect to the line of sight between the Sun and the respective spacecraft, and the bottom panels show the relative elevation to the ecliptic plane. The horizontal black dashed lines show the power threshold used to discriminate the measurements to be plotted. }

\label{fig2}
\end{figure}

During the period of observation STEREO A was at about $20.8^{\circ}$ of separation, in the Earth direction of rotation and at a distance from the Sun of 0.967~AU, while STEREO B had a separation of about $21.6^{\circ}$ in the opposite direction and a distance of 1.027~AU from the Sun.

As mentioned in the previous section, we characterize source directions with respect to either spacecraft in terms of an azimuth and an elevation.\footnote{The elevation is defined as the angle measured between the ecliptic plane and the source direction. \inlinecite{2009SoPh..259..255R} makes use of co-latitude instead of elevation, which has a simple relation that is defined as $\mathrm{elevation} = 90^{\circ} - \mathrm{colatitude}$} In this context, the azimuth is defined as the angle between the STEREO-Sun line and the source direction. Similarly, the elevation is defined as the angle between the ecliptic plane and the source. Figure~\ref{fig2} shows the direction finding calculations for various frequencies for each of the STEREO spacecraft and also shows that the source azimuth decreases with frequency from the perspective of STEREO A.  This parallax suggests sources distributed approximately along a line of sight from STEREO B with the higher frequency sources closer to the Sun so that they have a greater azimuth with respect to STEREO A.  Also, a clear time delay is observed in the top panels in Figure \ref{fig2}; the implications of this delay will be discussed later.  This figure shows the practicality of the eigenvalue decomposition method described in Section~\ref{S-EDM}.

The 425~kHz signature was chosen to compare our calculations to \inlinecite{2009SoPh..259..255R}, who use an iterative method to determine the locations of the radio sources, based on theoretical relations of the antenna power.  \inlinecite{cecconi2008} describe the theoretically deduced auto- and cross-correlations, and the angles indicating the direction of arrival of the radio emission. 

Table~\ref{table1} shows the results obtained by the two methods.  The average values of azimuth calculated around the maximum signal for the 7 December 2007 event are about $-1.7^{\circ}$ for STEREO A and $-0.2^{\circ}$ for STEREO B; the elevation angles are $-5.2^{\circ}$ and $-0.7^{\circ}$, respectively. These results are close to the values reported in \inlinecite{2009SoPh..259..255R} ($-4.5^{\circ}$ of azimuth in the reference frame of STEREO A and $+3.0^{\circ}$ in the frame of STEREO B).  Although there is a small discrepancy between our azimuths and those of \inlinecite{2009SoPh..259..255R}, the results appear to be consistent with a \textit{Wind}/WAVES observation that shows a source azimuth from a near-Earth perspective of $-0.5^{\circ}$, also reported by \inlinecite{2009SoPh..259..255R}.

\begin{table}[htb]
	\begin{tabular}{lccccc}
	\hline
  			& \multicolumn{2}{c}{STEREO A} & \multicolumn{2}{c}{STEREO B} & \textit{Wind}\\
	 		& LSF & EVD & LSF & EVD & LSF\\
	\hline
	 Azimuth 	& $-4.5^{\circ}$ & $-1.7^{\circ}$	& $+3.0^{\circ}$& $-0.2^{\circ}$ & $-0.5^{\circ}$ \\
	 Elevation	& $\approx 0^{\circ}$$_*$ & $-5.2^{\circ}$ & $\approx 0^{\circ}$$_*$ &$-0.7^{\circ}$ & \,--\,\\
	\hline
	\end{tabular}\\
	$_*$ Reported as ``close to zero."
	\caption{Comparative table of the values obtained for the 7 December 2007 type III event, using a least-squares fitting  (LSF) algorithm and the eigenvalue decomposition (EVD) algorithm.}
	\label{table1}
\end{table}

\begin{figure}[ht]
\centering
\includegraphics[width=0.96\columnwidth]{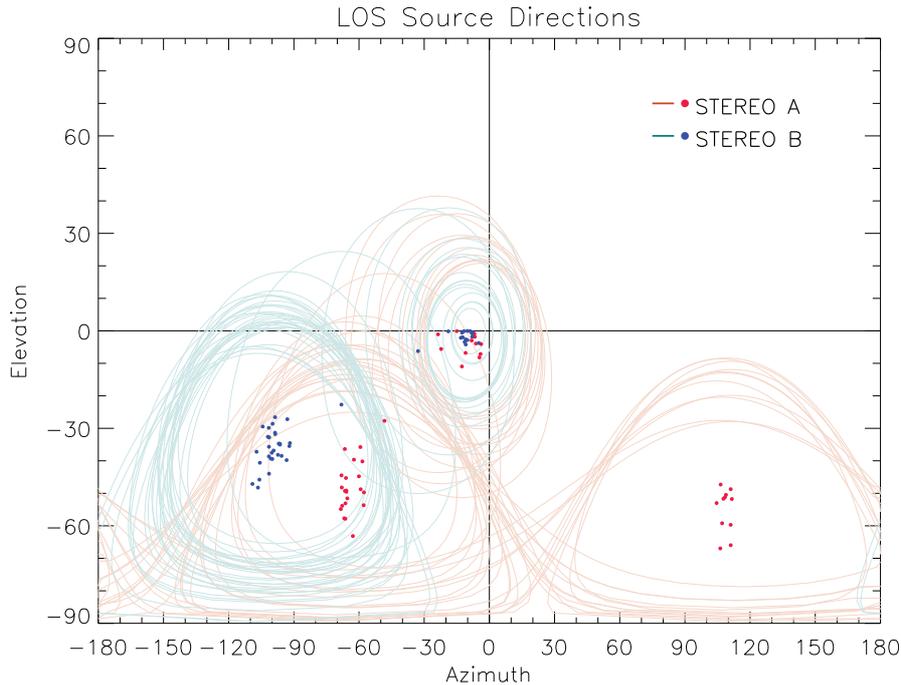}
\caption{Azimuth-elevation plot of the derived positions at 425~kHz for the 7 December 2007 solar type III burst. The concentration of points close to the origin represents the general location of the type III radio source. Data points signifying directions far from the Sun characterize pre- or post-flare instrumental and background noise. 
Also plotted for each such direction is a circle on the celestial sphere centered upon it whose radius is the characteristic radius, $r_\mathrm{c}$, of the respective location on which it is centered (see Appendix \ref{appendix-a}).  Note that circles on the celestial sphere are warped when projected into an azimuth-elevation format.}
\label{fig3}
\end{figure}

The positions of the 425~kHz radio sources relative to the Sun for each temporal observation are plotted in Figure~\ref{fig3}. The coordinates of the center of the solar disk are ($0^{\circ}$,$0^{\circ}$). The closed curves in Figure~\ref{fig3} represent circles on the celestial sphere centered upon respective source centroids and whose radii, $r_\mathrm{c}$, are the rms radii of a distribution of randomly phased sources so centered and consistent with the radio signatures (see Appendix \ref{appendix-a}).  Note that circles on the celestial sphere are generally warped when projected onto a planar azimuth-elevation format.  (Indeed, circles that separate the north and south poles project into curves that are not even closed in an azimuth-elevation mapping.)  This figure shows the source circles and their centroids for an extended period around the type III burst (04:30~UT to 05:00~UT), which include those for the background signal, which are centered in the southern hemisphere, much weaker, appear to be highly diffuse, and not necessarily the same for STEREO A as for STEREO B. 

The location of the source in the interplanetary medium was estimated by parallax, \textit{i.e.}, ``triangulation,'' based on \inlinecite{2010ApJ...710L..82L} using the mean values of line-of-sight azimuths and elevations plotted in Figure~\ref{fig2}. A brief summary of the trigonometry for this application is given in Appendix \ref{appendix-b}.

\begin{figure}[htp]
\centering
\includegraphics[width=0.8\columnwidth]{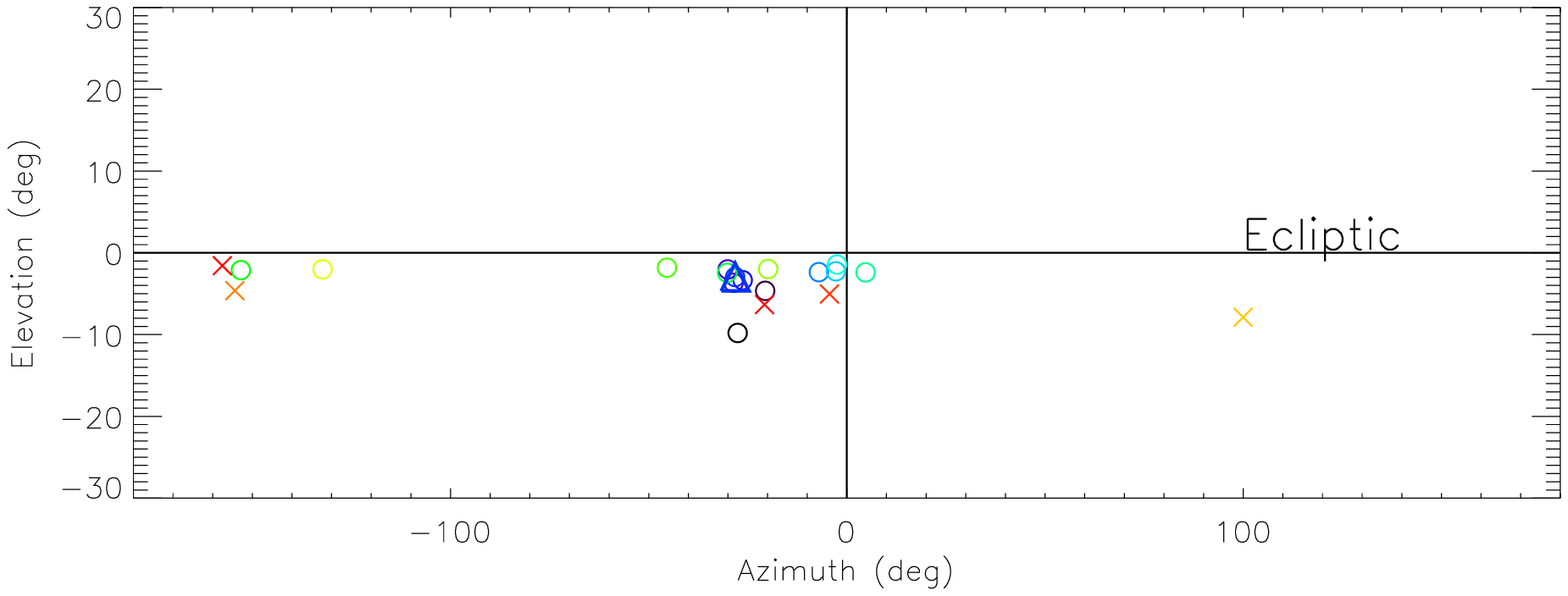}
\includegraphics[width=0.8\columnwidth]{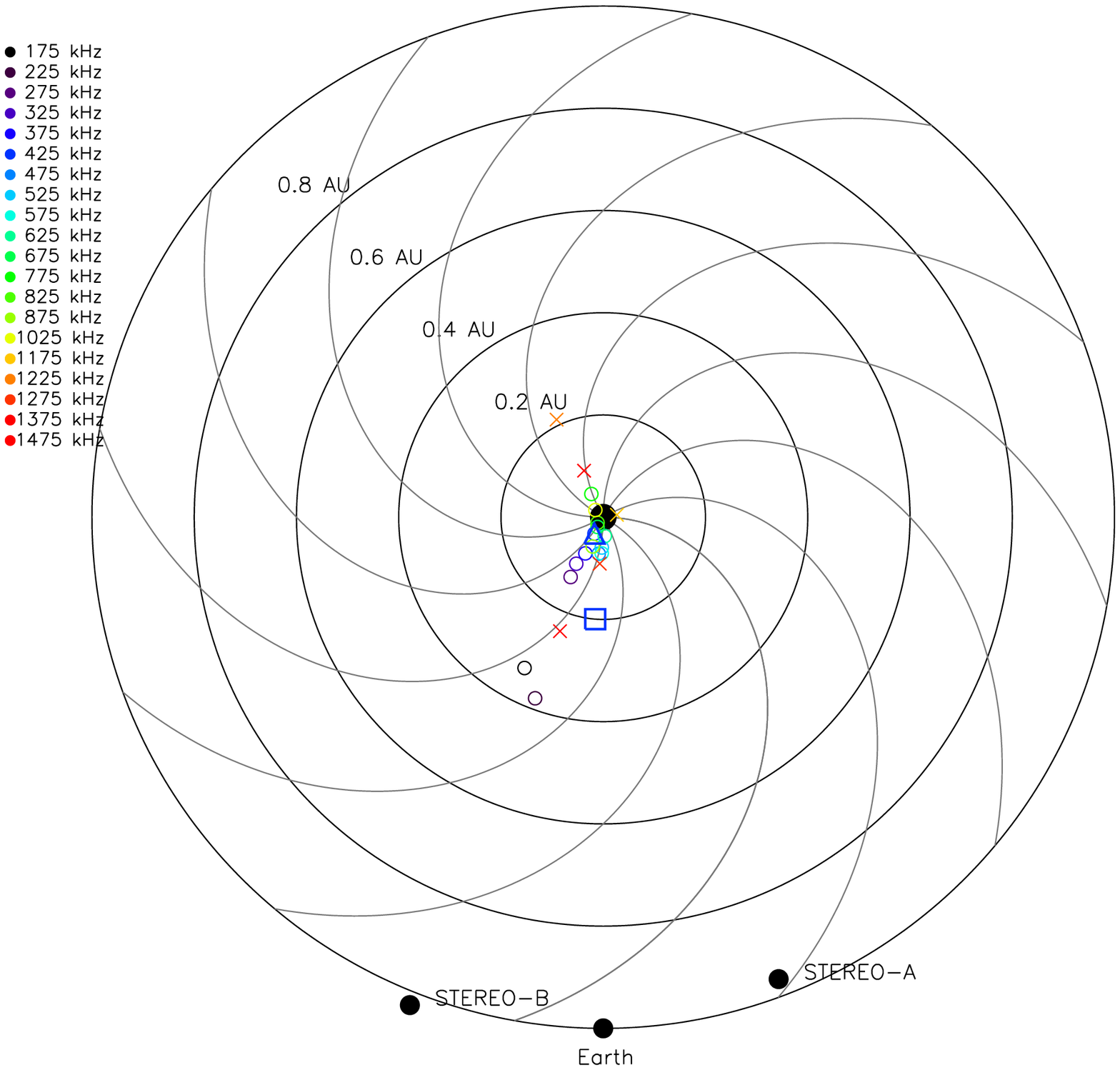}
\caption{Radio source locations are plotted in a Sun-centered reference frame with Earth at zero azimuth.  In this format, source azimuths are the Earth-Sun-source angle projected onto the ecliptic plane.  The horizontal axis in the top frame represents the ecliptic with the Sun at the origin. Bottom: Locations of the radio sources determined by parallax projected perpendicularly onto the ecliptic plane.  The 425~kHz source location is indicated by a triangle data point.  The location of the 425~kHz source reported by \protect \inlinecite{2009SoPh..259..255R} is indicated by a square data point.  The Parker spirals are plotted in gray and calculated using the formula $\phi = \phi_0 - (\Theta_{\odot}/V_{\mathrm{sw}}) r$, where $r$, is the radial distance to the Sun, $\mathrm \phi_0$ is an arbitrary angle, $\Theta_\odot$ is the rotational velocity of the Sun (2~km\,$s^{-1}$), and $\mathrm V_{\mathrm{sw}}$ is the solar wind velocity (400~km\,$s^{-1}$). The cross symbols represent positions that are unreliable due to uncertainties in the antenna pattern at high frequencies (see discussion at the end of Section~3.1)}
\label{fig4}
\end{figure}

The results of the parallax determination for the 7 December 2007 event are shown in Figure~\ref{fig4}; for the specific case of the 425~kHz frequency the sources are about $28^{\circ}$ east and $3^{\circ}$ south, at a distance of 0.04~AU (triangle), which is not consistent with the result found by \inlinecite{2009SoPh..259..255R}, of $2^{\circ}$ east and $\approx 0^{\circ}$ of elevation, at a heliocentric distance of 0.2~AU (square). This can easily be explained by the fact that a small change in the angular configuration may lead to large discrepancies in the parallax determination. The parallax algorithm works by finding the location of the intersection of the two vectors defined by the average positions. The fact that these locations are different, even by a small amount, means that the intersection point can be in a considerably different location, as demonstrated above. The algorithm was tested using the results from  \inlinecite{2009SoPh..259..255R}, obtaining values in the general neighborhood of those reported.  Using the electron density model described in  \inlinecite{1998SoPh..183..165L} with an electron density at 1~AU of $\mathrm{7.2~cm^{-3}}$, we calculated the distance where the radio emission could be produced for a range of frequencies. We found that, for this model, a wave of frequency 425~kHz is emitted at a distance of about 0.058 AU, which is within the limits of the values obtained by parallax, therefore showing the plausibility of our results, assuming that the radio emission was generated at the fundamental plasma frequency.\footnote{It should be noted that it is still unknown whether type III radio emission is produce at the fundamental or harmonic plasma frequency, or both.} 

Figure~\ref{fig4} also shows the canonical Parker spiral, representing the likely propagation path of the emitting electron beam generated by the flare. However, remember that electron beams have a physical size and are likely to be inhomogeneous. This,  along with changing conditions in the media into which they impinge, affects the location of the centroid of emission, which is what this direction finding technique and others want to determine. The frequency drift of the radio sources is seen in Figure~\ref{fig4};  high-frequency sources are located closer to the Sun, while low-frequency sources are further in interplanetary space. However,  bear in mind that  computed positions at higher frequencies may be unreliable due to uncertainties in the antenna pattern (see the discussion at the end of this section).

We now propose to examine the timing of the radio profiles at the two spacecraft as a prospective control resource for the locations determined by parallax.  We do this by first computing the distance to each spacecraft from the extrapolated locations. Then, times of flight for each spacecraft are computed assuming that the radio emission travels in a straight line from the source centroid to the spacecraft at a constant velocity, the speed of light.  The difference between these two times is compared with the time shift between radio flux profiles at the two spacecraft (see top panels of Figure~\ref{fig2}) to determine whether they are consistent with the source locations extrapolated by parallax.  This ``time-of-flight analysis'' assumes that the onsets of the signals at the two spacecraft are the signature of radio emission simultaneously emitted from a single compact source.  The limitations of this ``time-of-flight analysis'' are in the temporal resolution of the measurements and errors inherent in the assumptions of compactness and simultaneity\footnote{The real radio sources appear to be spatially extended, and emission from spatially extended sources can be directional, emanating toward one spacecraft at one time and the other at another time.}

We find that the time shift computed from the direction finding results ranges from $\approx$0.9 min at the lowest frequencies decreasing to $\approx$0.15 min at the highest, while the observed delay between the onsets of the emission received by STEREO A and  B range from 1.3 to 0.65 min, respectively.  For a comparison to other work, we consider the 425~kHz radio source.  For this frequency, we find a time delay of 0.4 min for both our location and that of \inlinecite{2009SoPh..259..255R}, although the former is significantly more distant than the latter.  The time delay between the radio signatures is about 0.65 min. We found that the direction finding and time-of-flight analysis results are consistent within the errors inherent to both techniques.

For frequencies higher than $\approx$1000~kHz the radio signatures are unreliable, the parallax algorithm sometimes giving obviously unrealistic locations.   The antenna pattern is more complicated at high frequencies, as the radio wavelengths approach the quarter-wave resonance of the antenna system \cite{Bale:2008p4}, making the antenna calibration more difficult. Ongoing efforts to improve the calibration of the STEREO/WAVES instrument based on a better understanding of the effective antenna lengths and directionality will improve the results of the direction finding techniques in the future at high frequencies.  The foregoing uncertainties do not necessarily imply that the parallax determinations are generally wrong, just that they are unreliable.

\subsection{29 January 2008 Type III Event}

\begin{figure}[ht]
\centering
\includegraphics[width=0.9\columnwidth]{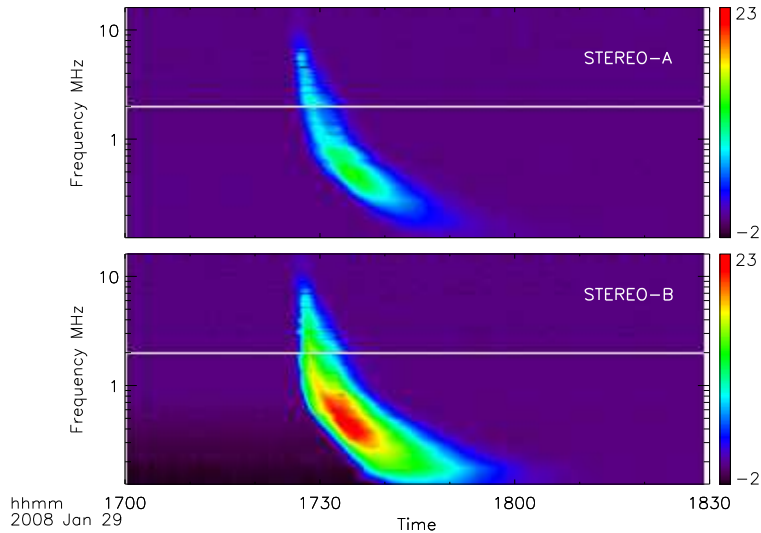}
\caption{STEREO A and  B dynamic spectra of the 29 January 2008 type III burst from 04:00~UT to 06:00~UT. The plotted frequencies range from 100~kHz to 16.025~MHz. The color bars represent the intensity of the radio emission measured in decibels.}
\label{fig5}
\end{figure}

The other type III event studied by \inlinecite{2009SoPh..259..255R} occurred on 29 January 2008. This specific type III was associated with a GOES B1.2 class flare,  hosted by the active region 10982, which was located close to the east limb of the Sun. The flare began at 17:28~UT, reaching maximum at 17:34~UT and ending at 17:43~UT (Figure~\ref{fig5}). The maximum of the radio  emission was  observed at 17:45~UT \cite{2009SoPh..259..255R} by both STEREO spacecraft.  On this date both STEREO spacecraft had a separation between them of $45.2^{\circ}$. STEREO A was located $21.6^{\circ}$  west, at a radial distance of 0.967~AU, while STEREO B was at $23.5^{\circ}$ to the east of the solar disk at a distance of 1.0015~AU.  

\begin{figure}[p]
\centering
\includegraphics[angle=90, clip,width=0.75\columnwidth]{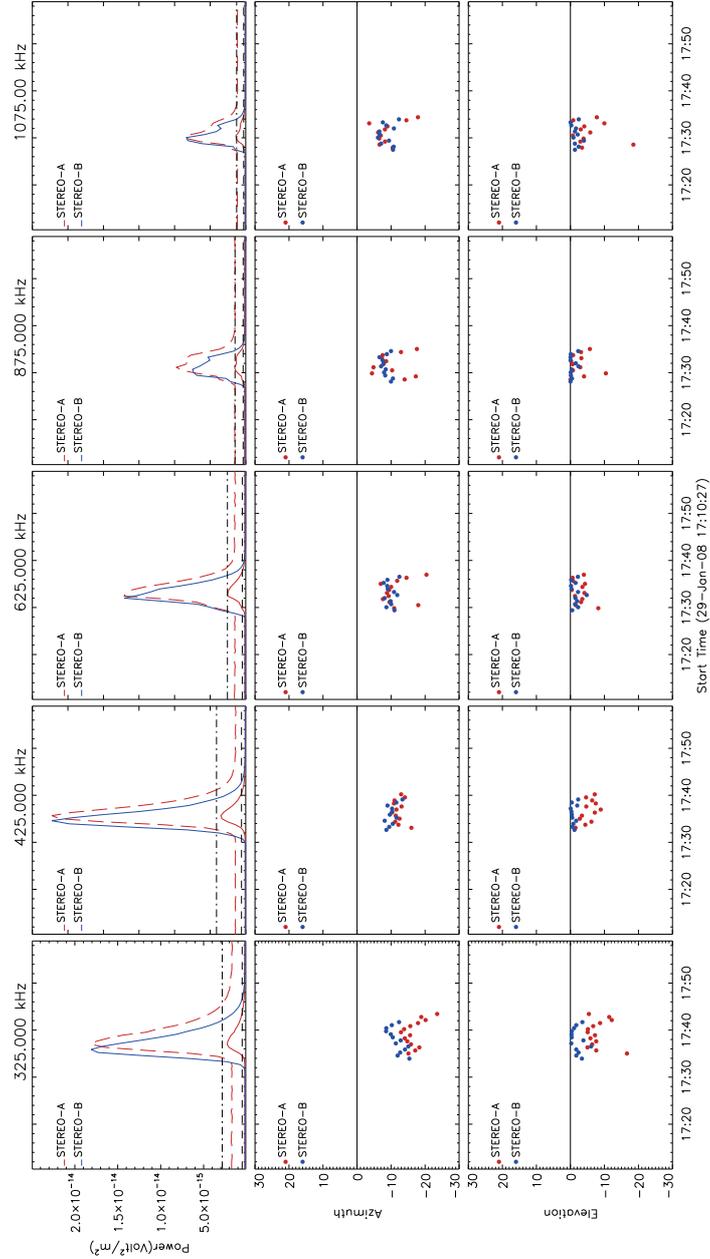}
\caption{Results of the direction finding analysis at STEREO A and B at 425, 625, 775, 925 and 1075~kHz for the 29 January 2008 solar type III burst. The top panels show the derived radio emission power, the middle panels show the derived azimuthal angles of the radio source and the bottom panels show the relative elevation to the ecliptic plane. The red dashed lines show the enhanced STEREO A power for a better visualization of the signal time delay between the two spacecraft. The horizontal black dashed lines show the power threshold used to plot only the azimuth and elevation associated with the type III burst.}
\label{fig6}
\end{figure}

Figure~\ref{fig6} shows the radio flux and source directions relative to both spacecraft.  The radio flux was an order of magnitude more intense at the location of STEREO B than at STEREO A.  There is also a greater spread of source directions for STEREO A.  For STEREO A, we find a mean azimuth around the maximum of power at 425~kHz of $-11.4^{\circ}$ and a mean elevation of about $-6.3^{\circ}$.  For STEREO B we find a mean azimuth of about  $-10.0^{\circ}$ and a mean elevation of  $-0.6^{\circ}$. These results are in reasonable agreement with \inlinecite{2009SoPh..259..255R}.  A comparison is shown in Table~\ref{table2}

\begin{table}[htb]
	\begin{tabular}{lccccc}
	\hline
  			& \multicolumn{2}{c}{STEREO A} & \multicolumn{2}{c}{STEREO B} & \textit{Wind}\\
	 		& LSF & EVD & LSF & EVD & LSF\\
	\hline
	 Azimuth 	& $-11.0^{\circ}$ 	& $-11.4^{\circ}$	&  $-8.5^{\circ}$ 	&  $-10.0^{\circ}$ 	& $-11.5^{\circ}$ \\
	 Elevation	& $\approx 0^{\circ}$$_*$ 	& $-6.3^{\circ}$ 	&  $\approx 0^{\circ}$$_*$ 	&$-0.6^{\circ}$ 		& \,--\,\\
	\hline
	\end{tabular} \\
	$_*$ Reported as ``close to zero."
	\caption{Comparative table of the values obtained for the 29 January 2008 type III event, using a LSF algorithm and the EVD algorithm.}
	\label{table2}
\end{table}

\begin{figure}[htb]
\centering
\includegraphics[width=0.80\columnwidth]{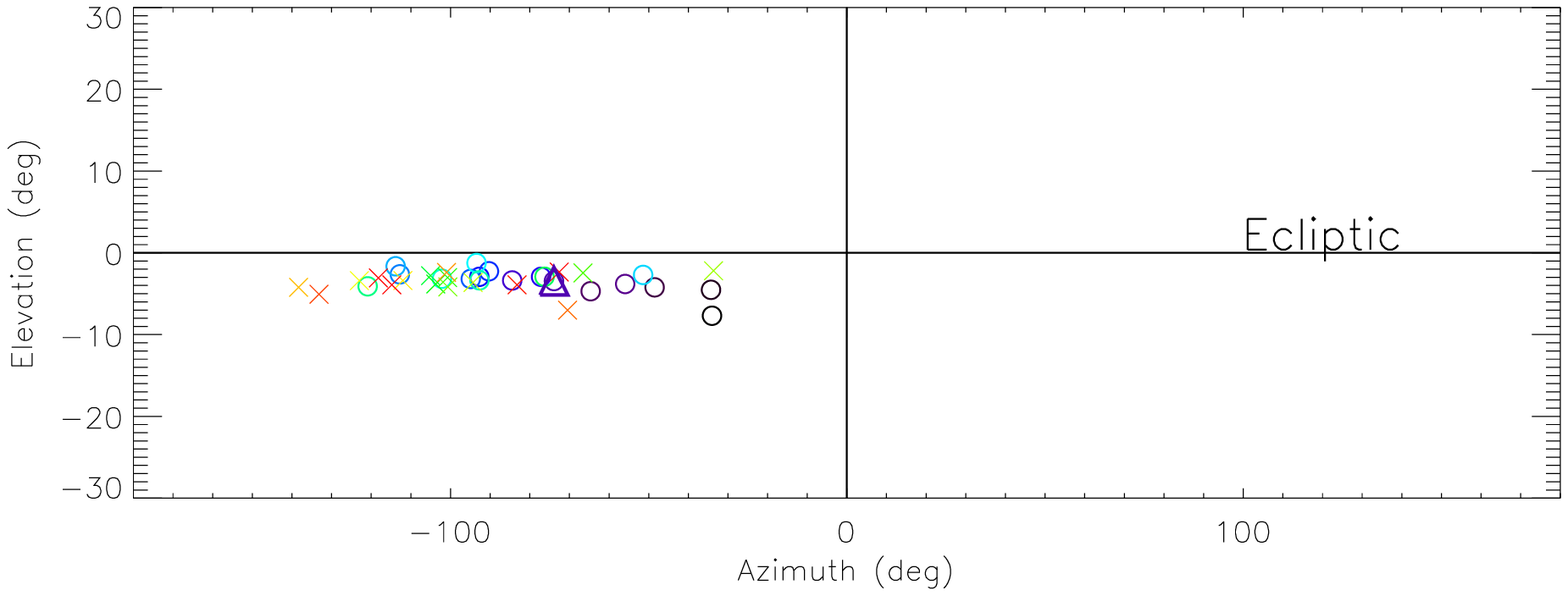}
\includegraphics[width=0.80\columnwidth]{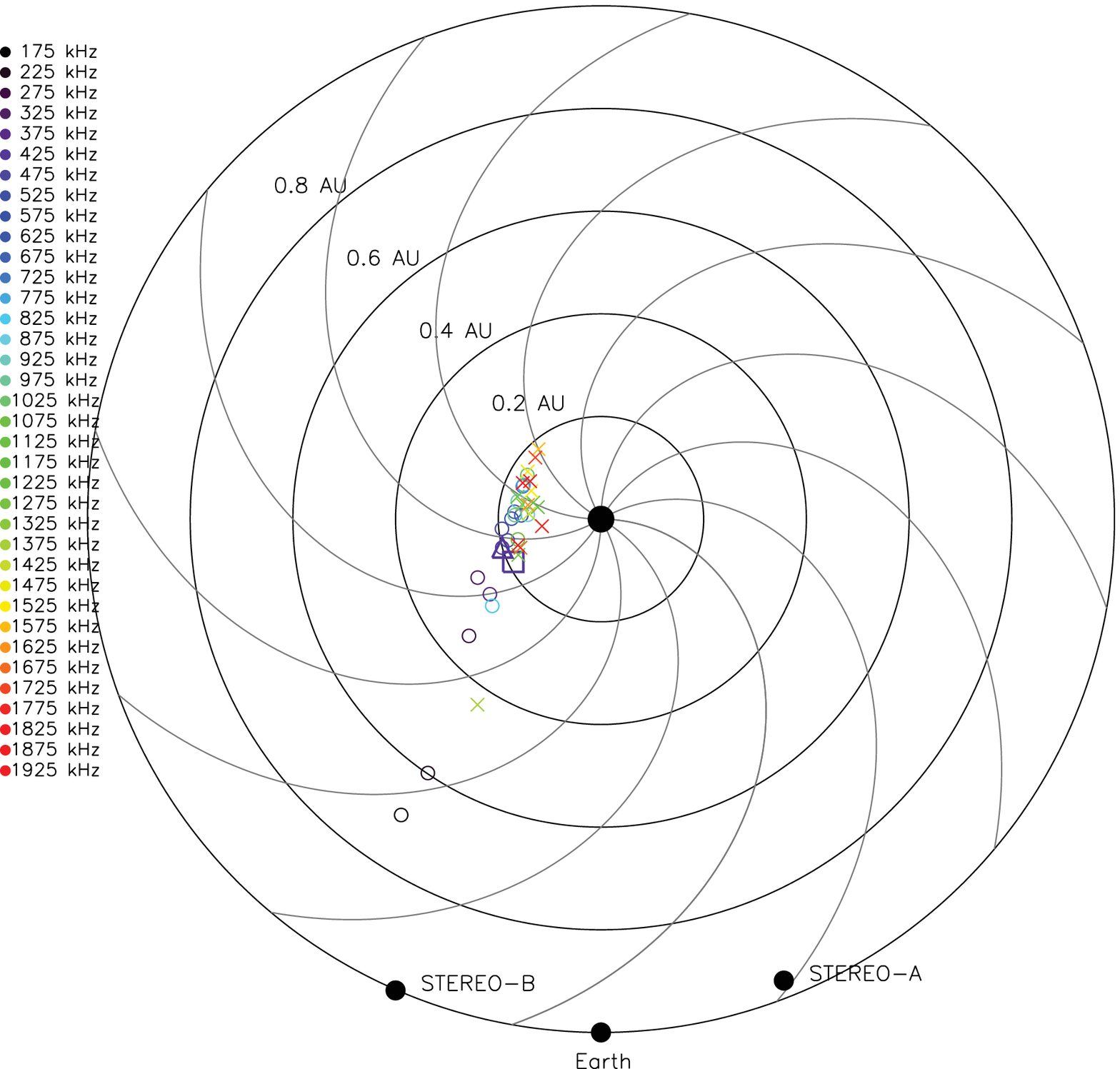}
\caption{Location of radio sources in the interplanetary space associated with the type III radio burst of 29 January 2008, for all operational frequencies.  The location we derive for the 425~kHz source is marked by a triangle.  The location of the same reported by \protect \inlinecite{2009SoPh..259..255R} is marked by a square.  The Parker spiral is plotted in gray and calculated assuming a constant velocity of 400~km~s$^{-1}$. Positions at high frequencies are expected to be incorrect (see Section~\ref{sec31}).  However, they are not obviously implausible, as seen for the event of 2007 December 07.}
\label{fig7}
\end{figure}

Figure~\ref{fig7} plots the source locations, determined by parallax. We found that the 425~kHz source was located at about $73.0^{\circ}$ east and at elevation of about $15.0^{\circ}$ south at a distance of approximately 0.21~AU (triangle). These results are almost in agreement with the results obtained by \inlinecite{2009SoPh..259..255R}, of $64.0^{\circ}$ east and a heliocentric distance of 0.19~AU (square in Figure~\ref{fig7}), given that slight differences in the angles used in the parallax calculation can lead to substantial discrepancies.  This gives a difference of about $9.0^{\circ}$ in azimuth and 0.02~AU in radial distance between our results and the results of \citeauthor{2009SoPh..259..255R}. In their work, \inlinecite{2009SoPh..259..255R} do not give precise information about the elevation angle obtained by parallax, although they describe that the location of the source is close to the ecliptic plane.

The time-of-flight analysis for this event using the direction-finding calculated data prescribes time delays from $\approx$1.7~min at the lowest frequencies decreasing to 0.45 min at the highest.  The radio profiles themselves show time delays from $\approx$1.4 min for the lowest frequencies decreasing to $\approx$0.0 at the highest (see Figure~\ref{fig6}: top panels).  For a frequency of 425~kHz, we expect a time delay of 1.0~min for our location, compared with 1.1~min for that of \inlinecite{2009SoPh..259..255R}.  The time delay in the radio flux is about 0.65 min, about the temporal sampling interval of the STEREO/WAVES instrument. The results of the time-of-flight analysis for the 29 January 2008 type III radio burst confirm the results obtained by the direction-finding algorithm, showing that the locations obtained are plausible.

\section{Conclusions}
     \label{S-Conclusions} 

The radio direction finding technique is a  powerful tool in the study of radio emissions in the interplanetary medium. Several different techniques  have been developed during the years to study type II and type III solar radio bursts (\opencite{1972Sci...178..743F}; \citeauthor{2009SoPh..259..255R}, \citeyear{1998JGR...10329651R,2009SoPh..259..255R}; \opencite{1995RaSc...30.1699L}; \opencite{2004JGRA..10909S17V}; \opencite{santolik2003}; \opencite{2005RaSc...40S3003C}; \opencite{2010RaSc...45S3003H}).  For three-axis-stabilized spacecraft, such as STEREO, some of these techniques have been applied with some success, leading to new insight into heliospheric radio emission during flares and CMEs.

We have developed a simple eigenvalue decomposition method to find the locations of radio sources based on observations from STEREO/WAVES.  This method can be easily adapted to observations from other missions, such as \textit{Solar Probe Plus} and \textit{Solar Orbiter}.  We have applied the method to two observational data sets already analyzed by \inlinecite{2009SoPh..259..255R} with similar results. For the 7 December 2007 type III event, the results of the algorithm are closer to the results obtained by the \textit{Wind} spacecraft, with a deviation of $\approx 3^{\circ}$ from the results reported by \inlinecite{2009SoPh..259..255R}. A slightly smaller deviation was found for the 29 January 2008 type III event. The statistical uncertainty in the source directions determined by both EVD and LSF techniques appears to be about $\approx 1.0^{\circ}$ \cite{cecconi2008,2010RaSc...45S3003H}.  The difference between the EVD and LSF source locations tends to be in this neighborhood, with significant exceptions. 

Based on \inlinecite{2010ApJ...710L..82L}, we used parallax to extrapolate the locations of the radio sources for both type III events studied by \inlinecite{2009SoPh..259..255R}, for all operative frequencies of the HFR1 receiver of the WAVES instrument.  When the parallax is small, \textit{i.e.}, the heliocentric angle between the spacecraft is narrow, small errors of the order of a few degrees in the angular position of the source introduce large uncertainties in the distance of the source from either spacecraft, and hence in its location with respect to the Sun.  Significant angular uncertainties are inherent in spatially extended source distribution, since these span wide angles that depend upon perspective.  Modeling of radio signatures resulting from realistic source distributions conditions could give us insight into discrepancies between the simple models based upon point sources and the reality.

A time-of-flight analysis was performed to check the direction finding solutions. In both cases, the trend in the time of flight is according to expectation for a type III burst emanating from a source region in the interplanetary medium above the east solar limb as viewed from Earth or either spacecraft.  The burst emission arrives at the spacecraft at different times within $\pm$2~min.  We attribute the onset-time difference to the two spacecraft being at different distances from the source, hence resulting in different travel times.  For both of the events we examined, the travel-time difference decreases with increasing frequency.  For the geometry implied by parallax, this is consistent with high-frequency sources been closer to the Sun than low-frequency sources.  For the 29 January 2008 event, the onset-time differences are somewhat less than those expected at the lowest and highest frequencies, assuming that the radio signals emanate simultaneously from the single point extrapolated by parallax and travel at the speed of light to the respective spacecraft.  The difference between the extrapolated and measured time delays appears to be related to the extended spatial distribution of the sources, wherein the radiation received by one spacecraft can have emanated from a significantly different part of the source region than the other.  Moreover, the lines of sight from STEREO A to the high-frequency sources pass closer to the Sun than the radial distances extrapolated for the source centroids.  This radiation is therefore subject to significant refraction, complicating the geometry of what part of the source distribution STEREO A sees and when it sees it.

\appendix
\section{Modeling Radio Source Distributions}
\label{appendix-a}

In general, it is widely accepted that type III radio bursts are only weakly polarized, with some special cases presenting a degree of circular polarization up to 25\% (\textit{i.e.}, \opencite{springerlink:10.1007/s11207-007-0277-8}). Ideally, then, the real correlation ellipsoid of the type III bursts should be an oblate spheroid, with the two major axes equal.Whether or not this is exactly the case, we propose to characterize type III bursts with a size parameter that supposes that this is the case. For that purpose, it is useful to consider a model in which the radiation received by the receiver is an incoherent superposition of sources uniformly distributed over a cone of radius $\theta_0$, and to determine the shape of the spheroid for such a model.We begin by considering a reference frame with the receiver at the origin and where the $z$-axis is the axis of the cone.We first consider a source in the $x$--$z$-plane and angle $\theta$ from the $z$-axis emitting radiation polarized in the $x$--$z$-plane of unit mean power at the receiver. In that case, 

\begin{eqnarray} 
\langle E_x^2\rangle &=& \cos^2{\theta},\nonumber\\ 
\langle E_y^2\rangle &=&0, {\rm ~and}\\
\langle E_z^2\rangle &=&\sin^2{\theta}.\nonumber
\end{eqnarray} 
In the alternative case of an emitter in the same location but polarized in the $y$--$z$-plane,
\begin{eqnarray} 
\langle E_x^2\rangle &=& 0,\nonumber\\
\langle E_y^2\rangle &=& 1, {\rm ~and}\\
\langle E_z^2\rangle &=& 0.\nonumber
\end{eqnarray} 
An unpolarized source at this location can be represented as an incoherent superposition of the above:
\begin{eqnarray} 
\langle E_x^2\rangle &=& \cos^2{\theta},\nonumber\\
\langle E_y^2\rangle &=& 1, {\rm ~and}\\
\langle E_z^2\rangle &=& \sin^2{\theta}.\nonumber
\end{eqnarray} 
For this type of source, we note that
\begin{equation}
\langle E_z^2\rangle = \sin^2{\theta} = r^2,
\end{equation}
where 
\begin{equation}
r \equiv \sin{\theta}
\end{equation}
is the radius of the circle on the unit sphere that makes an angle $\theta$ with the axis of the cone.
Moreover,
\begin{equation}
\langle E^2\rangle ~\equiv~ \langle E_x^2 + E_y^2 + E_z^2\rangle = 2.
 \end{equation}
 
For a distribution, $D(\Omega)$, of incoherent sources over the interior of a conical domain of solid angle, $\Omega$, whose half angle is $\theta_0$,
\begin{equation}
 \langle E_z^2\rangle = \int_{\theta < \theta_0} r^2 D(\Omega) d\Omega,
\end{equation}
while
\begin{equation}
 \langle E^2\rangle = 2\int_{\theta < \theta_0} D(\Omega) d\Omega.
\end{equation}
In other words,
\begin{equation}
{\langle E_z^2\rangle \over \langle E^2\rangle} = {\langle r^2\rangle \over 2},
\end{equation}
where $\langle r^2\rangle$ is the mean square radius of the distribution from the conical axis of points on the unit sphere for angles less than $\theta_0$.
Accordingly, we characterize the source distribution by the following characteristic radius from the centroid of the source distribution:
\begin{equation}
r_\mathrm{c} = \Big(2{\langle E_z^2\rangle \over \langle E^2\rangle}\Big)^{1/2}.
\end{equation}

\section{Ecliptic-Plane Perspective of Source Locations}
\label{appendix-b}

The method we use to represent source locations in an ecliptic-plane-based reference frame is taken from \inlinecite{2010ApJ...710L..82L} who developed the trigonometry we use and explained its limitations.  The method defines the source position in space in terms of the lines proceeding from the individual STEREO spacecraft in their respective directions to the supposedly single source.  The intersection of these lines projected onto the ecliptic plane defines the ecliptic azimuth, $\alpha$, of the source.  If $\alpha_\sigma$ is the azimuthal angle between the Sun center and STEREO A or B ($\sigma \in \{\textrm{A, B}\}$), $f$ representing the ratio between the heliographic distance to STEREO A ($d_\mathrm{A}$) and STEREO B ($d_\mathrm{B}$), and $\gamma$ the known longitudinal separation between the two spacecraft, then the angle, $\beta_\mathrm{A}$, between the source azimuth and that of the spacecraft, for example, can be expressed by
\begin{equation}
\tan \beta_\mathrm{A} = \left\{ 
  \begin{array}{l l}
    \frac{\sin \alpha_\mathrm{A} \sin(\alpha_\mathrm{B} + \gamma ) - f \sin \alpha_\mathrm{A} \sin \alpha_\mathrm{B}}{\sin \alpha_\mathrm{A} \cos(\alpha_\mathrm{B} + \gamma ) + f  \cos \alpha_\mathrm{A} \sin \alpha_\mathrm{B}} & \quad \mbox{if $ \alpha_\mathrm{A}  \geq 0$ and $ \alpha_\mathrm{B}  < 0$}\\
\\
    \frac{\sin \alpha_\mathrm{A} \sin(\alpha_\mathrm{B} - \gamma ) - f \sin \alpha_\mathrm{A} \sin \alpha_\mathrm{B}}{-\sin \alpha_\mathrm{A} \cos(\alpha_\mathrm{B} - \gamma ) + f  \cos \alpha_\mathrm{A} \sin \alpha_\mathrm{B}} & \quad \mbox{if $ \alpha_\mathrm{A}  \geq 0$ and $ \alpha_\mathrm{B}  \geq 0$}\\
\\
    \frac{\sin \alpha_\mathrm{A} \sin(\alpha_\mathrm{B} + \gamma ) - f \sin \alpha_\mathrm{A} \sin \alpha_\mathrm{B}}{-\sin \alpha_\mathrm{A} \cos(\alpha_\mathrm{B} + \gamma ) + f  \cos \alpha_\mathrm{A} \sin \alpha_\mathrm{B}} & \quad \mbox{if $ \alpha_\mathrm{A}  < 0$ and $ \alpha_\mathrm{B}  < 0$}
  \end{array} \right.
\end{equation}
Knowing the elevations, $e_\sigma$, of the source with respect to the ecliptic, it is then straightforward to determine the distance of the source above the ecliptic plane.  Errors in the determination of the azimuths and elevation can generally lead to different distances above the ecliptic plane insofar as the lines from the STEREO spacecraft toward the source fail to intersect.  In this study, we take the elevation of the source to be the mean value of the observed elevation angles.

\bibliographystyle{spr-mp-sola}
\bibliography{biblio} 

\end{article} 

\end{document}